\documentclass[a4paper,11pt,fleqn]{article}

\usepackage{amsmath,amsthm,amsthm,amsfonts,mathtools}
\usepackage[longnamesfirst]{natbib}
\usepackage{booktabs,bm,enumitem}
\usepackage[labelsep=period]{caption}
\usepackage[a4paper]{geometry}
\usepackage{setspace}
\usepackage[colorlinks=false, pdfborder={0 0 0}]{hyperref}
\usepackage{rotating}
\usepackage{cases}
\usepackage{authblk}
\usepackage{graphicx}
\usepackage{threeparttable}
\usepackage{pdflscape}
\usepackage{afterpage}
\usepackage{multirow}
\usepackage{xcolor}
\usepackage{titlesec}
\usepackage[titletoc,toc,title]{appendix}

\usepackage{xr}
\makeatletter
\newcommand*{\addFileDependency}[1]{% argument=file name and extension
  \typeout{(#1)}
  \@addtofilelist{#1}
  \IfFileExists{#1}{}{\typeout{No file #1.}}
}
\makeatother

\newcommand*{\myexternaldocument}[1]{%
    \externaldocument{#1}%
    \addFileDependency{#1.tex}%
    \addFileDependency{#1.aux}%
}
\myexternaldocument{supplement}

\allowdisplaybreaks

\usepackage[T1]{fontenc}

\numberwithin{equation}{section}

\theoremstyle{plain}

\theoremstyle{definition}

\renewcommand{\hat}{\widehat}

\title{On new tests for the Poisson distribution based on empirical weight functions}
\date{\today}
\author[1]{WC Kirui}
\author[1]{E Bothma}
\author[1]{M Smuts}
\author[2]{A Steyn}
\author[1]{IJH Visagie}

\affil[1]{Pure and Applied Analytics, North-West University, South Africa.}
\affil[2]{Centre for Business Mathematics and Informatics, North-West University, South Africa.}

\begin{document}

\date{\today}
\maketitle

\begin{abstract}

We propose new goodness-of-fit tests for the Poisson distribution. The testing procedure entails fitting a weighted Poisson distribution, which has the Poisson as a special case, to observed data. Based on sample data, we calculate an empirical weight function which is compared to its theoretical counterpart under the Poisson assumption. Weighted $L_p$ distances between these empirical and theoretical functions are proposed as test statistics and closed form expressions are derived for $L_1$, $L_2$ and $L_{\infty}$ distances. A Monte Carlo study is included in which the newly proposed tests are shown to be powerful when compared to existing tests, especially in the case of overdispersed alternatives. We demonstrate the use of the tests with two practical examples.

\vspace{0.5cm}

%\emph{MSC 2020 subject classifications: Primary 62D99, Secondary 62P20.}


\end{abstract}

\section{Introduction and motivation}
\label{Intro}

The Poisson distribution, originally introduced in \cite{poisson1828memoire}, is a useful model for count data with applications in various fields. For a detailed treatment of this distribution, together with its applications, see \cite{Hai1967}. The role of the Poisson distribution in biostatistics is explained in \cite{zar1999biostatistical}. For discussions of the properties and applications of the Poisson distribution, see \cite{patil1978weighted} and \cite{Johnson2005}. Several generalisations of this distribution have been proposed, for instance the generalised Poisson as well as the zero inflated Poisson distributions. An important generalisation, which plays a central role in this paper, is the weighted Poisson distribution introduced in \cite{fisher1934effect}.

Due to the wide range of applications of the Poisson distribution, it is often of practical interest to test the hypothesis that observed data are realised from a Poisson distribution. In order to proceed, we introduce some notation. Let $X_1,\dots,X_n$ be independent and identically distributed random variables with distribution function $F$.
%We denote the order statistics of the sample by $X_{(1)} \leq X_{(2)} \leq \dots \leq X_{(n)}$.
%The sample maximum is used extensively below, let $m = X_{(n)}$.
The Poisson distribution function, $F_{\lambda}$, with mean $\lambda > 0$, is
\begin{equation*}
    {F_{\lambda}(x) = e^{-\lambda}\sum_{k=0}^x \frac{\lambda^k}{k!}, \textrm{   for }  x \in \{0, 1, \dots \}}.
\end{equation*}
The corresponding probability mass function (pmf) is
\begin{equation*}
    {f_{\lambda}(x) = \frac{e^{-\lambda}\lambda^x}{x!}, \textrm{    for } x \in \{0, 1, \dots \}}.
\end{equation*}
Below we use the notation $P(\lambda)$ to indicate the Poisson distribution with mean $\lambda$.
We are interested in testing the composite goodness-of-fit hypothesis that
\begin{equation}
    {H_0 : F(x) = F_{\lambda}(x), \textrm{  for all } x \in \{0, 1, \dots \} \textrm{ and for some } \lambda > 0}.
\end{equation}
This hypothesis is to be tested against general alternatives. A recent review of existing tests for the Poisson distribution can be found in \cite{MV2020}, while a review of goodness-of-fit testing procedures for discrete distributions in general can be found in \cite{horn1977goodness} as well as \cite{kocherlakota1986goodness}.

The remainder of the paper is structured as follows. Section \ref{two} introduces new tests for the Poisson distribution and derives closed form expressions for the test statistics. Section \ref{three} shows a Monte Carlo power study, using a warp-speed bootstrap, in which the performances of the newly proposed tests are compared to that of existing tests. It is demonstrated that the new tests are competitive in terms of power. This section also includes two examples in which observed datasets are analysed and the hypothesis of the Poisson distribution is tested for each dataset. Finally, Section \ref{four} provides some conclusions.

\section{Newly proposed tests for the Poisson distribution}\label{two}

Below we introduce tests for the Poisson distribution based on an empirical version of the weight function of a weighted Poisson distribution. Before turning our attention to the construction of these tests, we consider the weighted Poisson distribution.

\subsection{The weighted Poisson distribution}

In \cite{fisher1934effect}, Fisher introduces the weighted Poisson distribution via the so-called method of ascertainment. \cite{Mij2020} points out that \cite{rao1965discrete} is often cited as the first paper to introduce the method of ascertainment, while \cite{fisher1934effect} introduces this method in a similar context three decades earlier. The idea underlying this method is based on a specified discrete random variable, meaning that the probability that a given value will realise is fixed beforehand. In certain practical situations, we might not be able to ``ascertain'' this value in some instances. That is, some of the values may not be observed and, therefore, go unnoticed. As a result, the probability of observing a specified value for the distribution is changed or re-weighted. This can be achieved by introducing a weight function giving more weight to the values which are likely to be ``ascertained'' and less weight to those which are not. This concept is made precise below.

Let $v$ be some function such that $v(x) \geq 0$ for $x \in \{0, 1, \dots\}$ and let $X \sim P(\lambda)$. $\widetilde{X}$ is said to be a weighted Poisson random variable, with parameter $\lambda$ and weight function $v$, if $\widetilde{X}$ has pmf
\begin{equation*}
    f_{\lambda,v}(x) = \frac{v(x)f_\lambda(x)}{E[v(X)]}, \text{  for  }x \in \{0, 1, \dots\},
\end{equation*}
where
\begin{equation*}
    E[v(X)] = \sum_{x=0}^{\infty} v(x)f_{\lambda}(x) < \infty.
\end{equation*}

When a constant weight function is used, the weighted Poisson distribution simplifies to the (unweighted) Poisson distribution. If $v(x)=c$, for some $c > 0$, then the pmf of $\widetilde{X}$ is
\[
    f_{\lambda,v}(x) = \frac{v(x)f_\lambda(x)}{E[v(X)]}=\frac{cf_\lambda(x)}{c}=f_\lambda(x).
\]
The above demonstrates that, given the value of $\lambda$, $v$ does not uniquely define $f_{\lambda,v}$ (since constant multiples of the weight function give rise to the same pmf). In order to ensure that, given $\lambda$, the chosen weight function uniquely determines the pmf, we define $w(x)=v(x)/E[v(X)]$. In this case, the weighted Poisson random variable, $\widetilde{X}$, has pmf.
\begin{equation}
    f_{\lambda,w}(x) = w(x)f_\lambda(x). \label{def of wp}
\end{equation}
Since, for every $v$ such that $E[v(X)]<\infty$, there exists a suitably rescaled weight function $w$, we take (\ref{def of wp}) to be the definition of the pmf of a weighted Poisson random variable.

In order to proceed, we introduce the empirical pmf:
\begin{eqnarray*}
 f_{n}(x) = \frac{1}{n} \sum_{j=1}^{n}I(X_{j}=x), \ \ x\in\{0,1,\dots\},
\end{eqnarray*}
with $I$ denoting the indicator function. In its most general form, $w$ can be chosen such that $f_{\lambda,w}$ takes the form of any specified pmf defined on the non-negative integers. As a result, given an observed dataset, $X_1,\dots , X_n$, we may estimate $\lambda$ by $\widehat{\lambda}=\sum_{j=1}^n{X_j}/n$ and then choose $w$ so that $f_{\widehat{\lambda},w}(x)=f_{n}(x)$ for all $x \in \{0, 1, \dots \}$.
%In order to ensure that $f_{n}(x) = f_{w}(x)$ for every non-negative integer, $w$ requires an infinite number of parameters.
Let $w^*$ denote the weight function equating $f_{\hat{\lambda},w^*}$ to $f_n$:
\begin{eqnarray} \label{wstar}
    w^*(x) = \frac{f_n(x)}{f_{\hat{\lambda}}(x)}.
\end{eqnarray}
We refer to $w^*$ as the empirical weight function.
%where $\hat{\lambda}$ is the maximum likelihood estimator of $\lambda$; i.e., the sample mean.
%We refer to $w^*$ as the empirical weight function.

Consider the case where the observed data are realised from a Poisson distribution. If the sample is sufficiently large, then we expect $w^*(x)$ to be approximately equal to $1$, at least for values of $x$ which are sufficiently likely to be observed. However, it should be noted that, for every $x$ exceeding the sample maximum, $w^*(x)=0$, see \eqref{wstar}. As a result, we construct tests for the Poisson distribution based on weighted $L_p$ distances between $w^*$ and the point $\mathbf{1}=\{1, 1, 1, \dots \}$, using weight functions which give more weight to smaller values of $x$.

%If the dataset under consideration is realised from a Poisson distribution, and sufficiently large, then we may expect that $w^*(x)$ would be approximately equal to $1$ for $x\in\{0,1,\dots,k\}$, where $k$ is some finite integer (depending on the sample size). As a result, we construct tests for the Poisson distribution based on weighted $L_p$ distances between $w^*$ and the point $\mathbf{1}=\{1, 1, 1, \dots \}$.
%note that both $w^*$ and the aforementioned point are in infinite dimensional space due to the unbounded support of the Poisson distribution.

\subsection{The proposed test statistics}

We now turn our attention to the calculation of weighted $L_p$ distances between the empirical weight function and the point $\mathbf{1}$ in order to construct test statistics. The sample maximum is used extensively below, we denote this by $m$. In the case where $p < \infty$, the use of weighted $L_p$ distances is required since the unweighted distance is infinite for every finite sample. The empirical pmf does not assign any probability weight to values larger than $m$, $f_n(x)=0$ for all $x>m$, meaning that $w^*(x)=0$ for all $x>m$. The distance of interest is
\begin{eqnarray*}
   L_p(w^*,\mathbf{1}) &=& \left(\sum_{x=0}^{\infty} |w^{*}(x)-1|^p\right)^{1/p} \\
   &=& \left( \sum_{x=0}^{m} |w^{*}(x)-1|^p + \sum_{x=m+1}^{\infty} |w^{*}(x)-1|^p\right)^{1/p} \\
   &=& \left(\sum_{x=0}^{m} |w^{*}(x)-1|^p + \sum_{x=m+1}^{\infty} 1\right)^{1/p}.
\end{eqnarray*}
Since the second summation above is infinite, we have that $L_p(w^*,\mathbf{1})=\infty$ for every finite sample. As a result, this distance cannot be employed as a test for the Poisson distribution. However, we may employ a weighted version of the $L_p$ distance between $w^*$ and $\mathbf{1}$, using some weight function $g$, such that $g(x)>0$ for all $x \in \{0,1, \cdots \}$ and $\sum_{x=0}^{\infty} g(x)<\infty$. Below we consider three choices of $g$, corresponding to the fitted Poisson pmf, $f_{\widehat\lambda}$, the empircal pmf, $f_n$, and a Laplace type weight function of the form $L(x)=\textrm{e}^{-ax}$, where $a>0$ is a user-defined tuning parameter.

Although the $L_\infty$ distance is finite, for a finite sample it would obtain a minimum value of $1$ by the same reasoning. As a result, we also opt to include a weight function when employing this distance as a test statistic.

Consider the weighted $L_{p,g}$ distance (where $g$ indicates the weight function used) between $w^*$ and $\mathbf{1}$ for $p<\infty$:
\begin{eqnarray}
L_{p,g}(w^*,\mathbf{1}) &=& \left( \sum_{x=0}^{\infty} |w^{*}(x)-1|^p g(x) \right)^{1/p} \nonumber \\
        &=& \left( \sum_{x=0}^{m} |w^{*}(x)-1|^p g(x) + \sum_{x=m+1}^{\infty} |w^{*}(x)-1|^p g(x) \right)^{1/p} \nonumber \\
        &=& \left( \sum_{x=0}^{m} |w^{*}(x)-1|^p g(x) + \sum_{x=m+1}^{\infty} g(x) \right)^{1/p}, \label{Tnpg}
\end{eqnarray}
which is finite since the first summation consists of a finite number of summands while, for the second summation, we have that \( \sum_{x=m+1}^{\infty}g(x) \leq \sum_{x=0}^{\infty}g(x)\) since $g(x) \geq 0$, and \(\sum_{x=0}^{\infty} g(x) < \infty \) by the definition of $g$.

The first three test statistics proposed are obtained by setting $p=1$ in (\ref{Tnpg}) together with the various choices of $g$ mentioned. In this case, (\ref{Tnpg}) simplifies to
\begin{equation}
    L_{1,g}(w^*,\mathbf{1}) = \sum_{x=0}^{m} |w^{*}(x)-1| g(x) + \sum_{x=m+1}^{\infty} g(x).
\end{equation}
Setting $g(x)=f_{\widehat{\lambda}}$, we obtain
\begin{eqnarray*}
    T_{n,f_{\widehat{\lambda}}}^{(1)}:=L_{1,f_{\widehat{\lambda}}}(w^*,\mathbf{1}) &=& \sum_{x=0}^{m} |w^{*}(x)-1| f_{\widehat{\lambda}}(x) + \sum_{x=m+1}^{\infty} f_{\widehat{\lambda}}(x)\\
    &=&\sum_{x=0}^{m} |w^{*}(x)-1| f_{\widehat{\lambda}}(x) + 1-F_{\hat{\lambda}}(m).
\end{eqnarray*}

The second proposed test statistic is obtained using the empirical pmf, $f_n$, as a weight function. In this case the test statistic can be expressed as
\begin{eqnarray*}
    T_{n,f_{n}}^{(1)}:=L_{1,f_{n}}(w^*,\mathbf{1}) &=& \sum_{x=0}^{m} |w^{*}(x)-1| f_{n}(x) + \sum_{x=m+1}^{\infty} f_{n}(x)\\
    &=&\sum_{x=0}^{m} |w^{*}(x)-1| f_{n}(x),
\end{eqnarray*}
where the final equality follows since $f_n(x)=0$ for all $x>m$.

The final weighted $L_1$ distance based test statistic considered is obtained using the Laplace type weight function, $L(x)=\textrm{e}^{-ax}$, for some $a>0$. Using a Monte Carlo study, we determined that the powers associated with these tests are remarkably insensitive to the choice of $a$. This insensitivity is also observed for tests based on the weighted $L_2$ and $L_{\infty}$ type distances. As a result, we set $a=1$ throughout the remainder of the paper, effectively reducing this weight function to $L(x)=\textrm{e}^{-x}$. The test statistic can, in this case, be expressed as
\begin{eqnarray*}
    T_{n,L}^{(1)}:=L_{1,L}(w^*,\mathbf{1}) &=& \sum_{x=0}^{m} |w^{*}(x)-1| \textrm{e}^{-x} + \sum_{x=m+1}^{\infty} \textrm{e}^{-x}\\
    &=&\sum_{x=0}^{m} |w^{*}(x)-1| \textrm{e}^{-x} + \frac{\textrm{e}^{-(m+1)}}{1- \textrm{e}^{-1}}.
\end{eqnarray*}

We now turn our attention to the tests based on weighted $L_2$ type distances. Using notation similar to that used above, we define the test statistics $T_{n,f_{\widehat{\lambda}}}^{(2)}$, $T_{n,f_{n}}^{(2)}$ and $T_{n,L}^{(2)}$. The required test statistics can be expressed as:
\begin{eqnarray*}
    T_{n,f_{\widehat{\lambda}}}^{(2)}&:=&L_{2,f_{\widehat{\lambda}}}(w^*,\mathbf{1}) =\left(\sum_{x=0}^{m} |w^{*}(x)-1|^2 f_{\widehat{\lambda}}(x) + 1-F_{\hat{\lambda}}(m)\right)^{1/2}, \\
    T_{n,f_{n}}^{(2)}&:=&L_{2,f_{n}}(w^*,\mathbf{1}) =\left(\sum_{x=0}^{m} |w^{*}(x)-1|^2 f_{n}(x)\right)^{1/2}, \\
    T_{n,L}^{(2)}&:=&L_{2,L}(w^*,\mathbf{1}) = \left(\sum_{x=0}^{m} |w^{*}(x)-1|^2 \textrm{e}^{-x} + \frac{\textrm{e}^{-(m+1)}}{1- \textrm{e}^{-1}}\right)^{1/2}.
\end{eqnarray*}

Next, we consider test statistics based on weighted $L_{\infty}$ distances. In general, we have that
\begin{eqnarray}
T_{n,g}^{(\infty)} &:=& L_{\infty,g}(w^*,\mathbf{1}) \\ &=& \underset{x \in \{0,1,\dots \}}{\textrm{max}} \left\{ |w^{*}(x)-1|g(x) \right\} \nonumber \\
&=& \textrm{max}\left\{\underset{x \in \{0,1,\dots,m \}}{\textrm{max}} \left\{ |w^{*}(x)-1|g(x) \right\},\underset{x \in \{m+1,m+2,\dots\}}{\textrm{max}} \left\{ |w^{*}(x)-1|g(x) \right\} \right\} \nonumber \\
&=& \textrm{max}\left\{\underset{x \in \{0,1,\dots,m \}}{\textrm{max}} \left\{ |w^{*}(x)-1|g(x) \right\},\underset{x \in \{m+1,m+2,\dots\}}{\textrm{max}} g(x) \right\}, \label{infgenform}
\end{eqnarray}
where the final equality follows from the fact that $w^*(x)=0$ for all $x$ greater than the sample maximum. Consider the final term of \eqref{infgenform}:
\begin{equation}
    \underset{x \in \{m+1,m+2,\dots\}}{\textrm{max}} g(x). \label{term}
\end{equation}
If $g(x)=f_n(x)$ is used as weight function, then this term can be omitted since $f_n(x)=0$ for all $x>m$. Since $g(x)=\textrm{e}^{-x}$ is a decreasing function of $x$, the term in \eqref{term} can be replaced by $\textrm{e}^{-(m+1)}$. In the case where $g(x)=f_{\widehat{\lambda}}(x)$, it can be shown that
\begin{equation*}
    \underset{x \in \{m+1,m+2,\dots\}}{\textrm{max}} f_{\widehat{\lambda}}(x) = f_{\widehat{\lambda}}(m+1),
\end{equation*}
for a derivation, see Appendix A.

The three test statistics based on weighted $L_{\infty}$ distances can be expressed as:
\begin{eqnarray*}
    T_{n,f_{\widehat{\lambda}}}^{(\infty)}&:=&L_{\infty,f_{\widehat{\lambda}}}(w^*,\mathbf{1}) =\textrm{max}\left\{\underset{x \in \{0,1,\dots,m \}}{\textrm{max}} \left\{ |w^{*}(x)-1|f_{\widehat{\lambda}}(x) \right\}, f_{\widehat{\lambda}}(m+1) \right\}, \\
    T_{n,f_{n}}^{(\infty)}&:=&L_{\infty,f_{n}}(w^*,\mathbf{1}) = \underset{x \in \{0,1,\dots,m \}}{\textrm{max}} \left\{ |w^{*}(x)-1|f_n(x) \right\}, \\
    T_{n,L}^{(\infty)}&:=&L_{\infty,L}(w^*,\mathbf{1}) = \textrm{max}\left\{\underset{x \in \{0,1,\dots,m \}}{\textrm{max}} \left\{ |w^{*}(x)-1|\textrm{e}^{-x} \right\},\textrm{e}^{-(m+1)} \right\}.
\end{eqnarray*}

\section{Numerical results}\label{three}

Below we compare the performance of the newly proposed tests to that of existing tests for the Poisson distribution. This is achieved through a Monte Carlo study in which empirical powers are calculated using a warp-speed bootstrap approach. Thereafter, we turn our attention to two observed datasets which have been modelled using a Poisson distribution and we demonstrate the use of the proposed techniques in order to test this assumption.

\subsection{Monte Carlo setup}

The finite sample powers below are calculated based on a significance level of $5\%$. We include results pertaining to sample sizes of 30, 50, and 100. We consider the performance of various tests against a range of alternative distributions. The pmf and notation used for each of the alternatives distributions used can be found in Table \ref{Alt dists}. These alternatives are selected since they are commonly used when testing the assumption of the Poisson distribution, see \cite{MV2020}, \cite{gurtler2000recent} and \cite{karlis2000simulation}.

\begin{table}[!htbp] \centering 
  \caption{Alternative distributions considered.} 
  \label{Alt dists} 
\begin{tabular}{@{\extracolsep{5pt}} lll} 
\\[-1.8ex]\hline 
Alternative distribution & Notation & Probability mass function  \\
\hline \\[-1.8ex] 
Discrete uniform & $DU(a,b)$ & $(b-a+1)^{-1}$  \\[1ex]
Binomial & $Bin(m,p)$ & ${m \choose x}p^{x}\left(1-p\right)^{x}$  \\[1ex]
Negative binomial & $NB(r,p)$ & ${{r+x-1} \choose x}p^{r}\left(1-p\right)^{x}$  \\[1ex]
Poisson Mixtures & $PM(p,\lambda_1,\lambda_2)$ & $(x!)^{-1}\left\{p\lambda_{1}^{x}\textrm{e}^{-\lambda_{1}}+\left(1-p\right)\lambda_{2}^{x}\textrm{e}^{-\lambda_{2}}\right\}$  \\[1ex]
Zero inflated Poisson & $ZIP(p,\lambda)$ & 
$\left(p\frac{x!}{e^{-\lambda}\lambda^{x}}\textrm{I}\left(x=0\right)+1-p\right)\frac{e^{-\lambda}\lambda^{x}}{x!}$  \\[1ex]
Weighted Poisson & $WP(\lambda,a,b)$ & $(y!)^{-1}\lambda^y\textrm{exp}(-\lambda) \frac{ay^2+by+1}{a(\lambda+\lambda^2) + b\lambda + 1}$ \\[1ex]
\hline \\[-1.8ex]
\end{tabular}
\end{table}

\subsection{Existing tests for the Poisson distribution}

We consider four tests based on the empirical distribution function (edf), 
\begin{equation*}
    F_n\left(x\right)=\frac{1}{n}\sum_{j=1}^n \textrm{I}(X_j \le x).
\end{equation*}
These tests include the classical Kolmogorov-Smirnov test:
\begin{equation*}
KS_n=\max_{x \in \{0,1,\dots \} }\left|F_{\widehat{\lambda}}\left(x\right)-F_n\left(x\right)\right|,
\end{equation*}
the Cram\'er-von Mises test:
\begin{equation}
    CV_n=\frac{1}{n}\sum_{x=0}^{\infty}(F_{\widehat{\lambda}}(x)-F_n(x))^{2}f_{\widehat{\lambda}}(x), \label{eq:CVM1}
\end{equation}
as well as the Anderson-Darling test:
\begin{equation}
AD_n=\frac{1}{n}\sum_{x=0}^{\infty}\frac{(F_{\widehat{\lambda}}(x)-F_n(x))^{2}f_{\widehat{\lambda}}(x)}{F_{\widehat{\lambda}}\left(x\right)\left(1-F_{\widehat{\lambda}}\left(x\right)\right)}. \label{eq:CVM2}
\end{equation}
Note that the calculation of $CV_n$ in \eqref{eq:CVM1} and $AD_n$ in \eqref{eq:CVM2} require the computation of an infinite summation. In both cases we use an approximation obtained by truncating the sum at $x = 100$.
The final edf based test considered is that proposed in \cite{klar1999goodness}. The corresponding test statistic is the sum of the absolute differences between the fitted and empirical distribution functions:
\begin{equation}
KL_n=\sqrt{n}\sum_{j=1}^{n}\left|F\left(X_{(j)}\right)-F_{\widehat{\lambda}}\left(X_{(j)}\right)\right|,\label{eq:L1 Norm}
\end{equation}
where $X_{(1)}, \dots, X_{(n)}$ denotes the order statistics of the sample.

\cite{klar1999goodness} also introduces a test statistic based on the supremum difference between the integrated distribution function (idf) of the Poisson distribution and the empirical version of this function. The idf is defined to be 
\begin{equation*}
    \Psi\left(t\right)=\int_{t}^{\infty}\left(1-F\left(x\right)\right)\textrm{d}x, \label{idf}
\end{equation*}
while the empirical idf is
\begin{equation*}
     \Psi_{n}\left(t\right)=\frac{1}{n}\sum_{j=1}^{n}\left(X_{j}-t\right)I{\left(X_{j}>t\right)}. \label{eidf}
\end{equation*}
The associated test statistic is denoted by
\begin{equation*}
    ID_n = \textrm{sup}_{t\geq0}\sqrt{n}\left|\Psi_{\widehat{\lambda}}\left(t\right)-\Psi_{n}\left(t\right)\right|.\label{eq:IDF}
\end{equation*}
Each of the tests discussed, rejects the null hypothesis for large values of the test statistic.

\subsection{Power calculations}

Since $\lambda$ is an unknown shape parameter, we use a parametric bootstrap procedure in order to approximate the distribution of the test statistics used. In order to speed up the required calculations we use the so-called warp-speed bootstrap, detailed in \cite{GPW:2013}, in order to arrive at the finite sample results shown. We use the algorithm below to implement the warp-speed bootstrap, this algorithm is an adapted version of the algorithm found in \cite{ABEV2020} and \cite{MV2020}.

\begin{enumerate}
\item Sample $X_{1}, \dots, X_{n}$ from distribution function $F$ and estimate $\lambda$ by $\widehat{\lambda}=\frac{1}{n} \sum_{j=1}^n X_j$.

\item Calculate the value of the test statistic: $S := S(X_1, \dots, X_n)$.

\item Generate $X_{1}^{\ast}, \dots, X_{n}^{\ast}$ from a $P(\widehat{\lambda})$ distribution. Calculate the test statistic based on this sample: $S^{\ast} = S(X_{n, 1}^{\ast}, \dots, X_{n, n}^{\ast})$.

\item Repeat Steps 1 to 3 $M$ times. Let $S_{m}$ be the value of the test statistic calculated using the $m^{\text{th}}$ dataset generated from $F$ and let $S_{m}^{\ast}$ be the value of the test statistic calculated from the bootstrap sample obtained in the $m^{\text{th}}$ iteration of the Monte Carlo simulation. As a result, we obtain $S_{1}, \dots, S_{M}$ and $S_{1}^{\ast}, \dots , S_{M}^{\ast}$.

\item We reject the hypothesis of the Poisson distribution for the $j^{\textrm{th}}$ sample from $F$ if $S_{j} > S_{\left( \left \lfloor M \cdot (1 - \alpha) \right \rfloor \right)}^{\ast}$, $j = 1, \dots, M$, where $S_{(1)}^{\ast} \leq \dotso \leq S_{(M)}^{\ast}$ are the order statistics obtained from the bootstrap samples.
\end{enumerate}

Table \ref{n50} shows the empirical powers obtained by the various tests considered for samples of size 50, the table shows the percentage of samples (rounded to the nearest integer) that results in a rejection of the null hypothesis against each of the alternative distributions considered. The results presented are based on 50 000 Monte Carlo replications. When discussing the results, we compare the powers achieved by the newly proposed tests and to those of the existing tests. Note that Table \ref{n50} contains a column indicating $FI$, the Fisher index of the alternative distribution considered. The discussion further distinguishes between alternatives based on their Fisher index. That is, we comment on performance against equidispersed as well as under and over dispersed alternatives. For ease of comparison, the highest power achieved against each alternative distribution is printed in bold. In the event that the maximal power is achieved by multiple tests, all of these instances are printed in bold font.
%Table 2 contains the empirical powers achieved by the tests for a sample of size 50. We discuss these powers in detail. Appendix A contains the powers achieved for samples of size 30, 100 and 200.

\begin{landscape}
\begin{table}[!htbp] \centering 
  \caption{Empirical powers obtained for samples of size n = 50} 
  \label{n50} 
  \footnotesize
\begin{tabular}{@{\extracolsep{5pt}} lccccccccccccccc} 
\\[-1.8ex]\hline 
\hline \\[-1.8ex] 
Distribution & FI & $KS_n$ & $CV_n$ & $AD_n$ & $KL_n$ & $ID_n$ & $T_{n,f_{\widehat{\lambda}}}^{(1)}$ & $T_{n,f_{n}}^{(1)}$ & $T_{n,L}^{(1)}$ & $T_{n,f_{\widehat{\lambda}}}^{(2)}$ & $T_{n,f_{n}}^{(2)}$ & $T_{n,L}^{(2)}$ & $T_{n,f_{\widehat{\lambda}}}^{(\infty)}$ & $T_{n,f_{n}}^{(\infty)}$ & $T_{n,L}^{(\infty)}$ \\
\hline
$P(0.5)$ & $1.00$ & $5$ & $5$ & $5$ & $4$ & $4$ & $5$ & $5$ & $5$ & $5$ & $5$ & $5$ & $5$ & $5$ & $4$ \\ 
$P(1)$ & $1.00$ & $5$ & $5$ & $5$ & $5$ & $5$ & $5$ & $5$ & $5$ & $5$ & $5$ & $5$ & $5$ & $5$ & $4$ \\ 
$P(5)$ & $1.00$ & $5$ & $5$ & $5$ & $5$ & $5$ & $5$ & $5$ & $4$ & $5$ & $5$ & $4$ & $5$ & $5$ & $4$ \\ 
$P(10)$ & $1.00$ & $5$ & $5$ & $5$ & $5$ & $5$ & $5$ & $5$ & $5$ & $5$ & $5$ & $5$ & $5$ & $5$ & $5$ \\ 
$DU(4)$ & $1.00$ & $45$ & $54$ & $\mathbf{63}$ & $60$ & $17$ & $62$ & $39$ & $34$ & $33$ & $0$ & $28$ & $43$ & $38$ & $28$ \\ 
$B(5,0.25)$ & $0.75$ & $18$ & $17$ & $18$ & $18$ & $\mathbf{23}$ & $13$ & $1$ & $15$ & $1$ & $0$ & $2$ & $12$ & $4$ & $17$ \\ 
$B(5,0.2)$ & $0.80$ & $14$ & $13$ & $13$ & $12$ & $\mathbf{15}$ & $10$ & $2$ & $11$ & $1$ & $0$ & $0$ & $11$ & $4$ & $13$ \\ 
$B(10,0.2)$ & $0.80$ & $10$ & $11$ & $11$ & $11$ & $\mathbf{15}$ & $7$ & $1$ & $7$ & $1$ & $0$ & $5$ & $8$ & $2$ & $8$ \\ 
$B(10,0.1)$ & $0.90$ & $7$ & $\mathbf{7}$ & $6$ & $6$ & $\mathbf{7}$ & $6$ & $2$ & $6$ & $1$ & $1$ & $2$ & $\mathbf{7}$ & $3$ & $6$ \\
$NB(9,0.9)$ & $1.11$ & $7$ & $7$ & $8$ & $8$ & $8$ & $6$ & $\mathbf{11}$ & $7$ & $\mathbf{11}$ & $\mathbf{11}$ & $\mathbf{11}$ & $5$ & $10$ & $6$ \\ 
$NB(45,0.9)$ & $1.11$ & $7$ & $6$ & $8$ & $9$ & $8$ & $6$ & $\mathbf{11}$ & $7$ & $10$ & $10$ & $7$ & $4$ & $10$ & $7$ \\ 
$PM(0.5,3,5)$ & $1.25$ & $13$ & $13$ & $19$ & $20$ & $\mathbf{21}$ & $9$ & $20$ & $13$ & $19$ & $16$ & $13$ & $5$ & $16$ & $13$ \\ 
$ZIP(0.9,3)$ & $1.30$ & $32$ & $21$ & $36$ & $31$ & $30$ & $28$ & $30$ & $\mathbf{57}$ & $24$ & $11$ & $56$ & $14$ & $20$ & $56$ \\ 
$PM(0.1,1,5)$ & $1.31$ & $20$ & $19$ & $31$ & $32$ & $32$ & $16$ & $31$ & $\mathbf{53}$ & $28$ & $15$ & $53$ & $6$ & $24$ & $52$ \\ 
$NB(15,0.75)$ & $1.33$ & $16$ & $17$ & $27$ & $28$ & $\mathbf{30}$ & $11$ & $29$ & $14$ & $28$ & $24$ & $13$ & $5$ & $23$ & $13$ \\ 
$NB(3,0.75)$ & $1.33$ & $21$ & $19$ & $25$ & $26$ & $27$ & $17$ & $\mathbf{32}$ & $24$ & $31$ & $28$ & $28$ & $11$ & $26$ & $21$ \\ 
$DU(6)$ & $1.33$ & $73$ & $78$ & $\mathbf{86}$ & $\mathbf{86}$ & $66$ & $78$ & $71$ & $80$ & $64$ & $5$ & $82$ & $44$ & $54$ & $84$ \\ 
$NB(4,0.7)$ & $1.43$ & $27$ & $25$ & $36$ & $37$ & $40$ & $20$ & $\mathbf{42}$ & $25$ & $40$ & $36$ & $37$ & $11$ & $34$ & $26$ \\ 
$NB(2,2/3)$ & $1.50$ & $37$ & $34$ & $42$ & $42$ & $44$ & $30$ & $\mathbf{48}$ & $40$ & $46$ & $42$ & $43$ & $19$ & $39$ & $35$ \\ 
$NB(3,2/3)$ & $1.50$ & $35$ & $32$ & $43$ & $44$ & $47$ & $26$ & $\mathbf{48}$ & $33$ & $46$ & $41$ & $46$ & $15$ & $40$ & $32$ \\ 
$ZIP(0.8,3)$ & $1.60$ & $86$ & $67$ & $84$ & $79$ & $78$ & $78$ & $78$ & $\mathbf{95}$ & $71$ & $26$ & $94$ & $64$ & $69$ & $\mathbf{95}$ \\ 
$PM(0.2,1,5)$ & $1.61$ & $58$ & $55$ & $75$ & $75$ & $75$ & $47$ & $67$ & $\mathbf{81}$ & $63$ & $35$ & $\mathbf{81}$ & $17$ & $52$ & $79$ \\ 
$NB(1,0.5)$ & $2.00$ & $75$ & $74$ & $80$ & $77$ & $79$ & $66$ & $\mathbf{82}$ & $78$ & $79$ & $73$ & $77$ & $51$ & $71$ & $70$ \\ 
\hline \\[-1.8ex] 
\end{tabular} 
\end{table} 
\end{landscape}

The results in Table \ref{n50} indicate that all tests considered achieve the specified nominal significance level of 5\% closely. Equidispersed alternatives to the Poisson are not particularly prevalent in the statistical literature, as a result, we include a single equidispersed alternative: $D[0,4]$. The $AD_n$ test performs best against this alternative, achieving an empirical power of 63\% for a sample of size 50. This performance is closely followed by $T_{n,f_{\widehat{\lambda}}}^{(1)}$ which achieves a power of 62\%.

When turning our attention to the underdispersed alternatives, many of the newly proposed tests do not perform well. Notable exceptions are $T_{n,f_{\widehat{\lambda}}}^{(1)}$, $T_{n,f_{\widehat{\lambda}}}^{(\infty)}$ and $T_{n,L}^{(\infty)}$, which exhibit powers that are, in some instances, comparable to those of the $ID_n$ test (the best performing test against these alternatives).

The final class of alternatives considered is the overdispersed distributions. The newly proposed tests notably outperform the existing tests for the Poisson distribution against the majority of the overdispersed alternatives considered, $T_{n,f_{n}}^{(1)}$ outperforms all other tests in 7 out of 14 instances considered. Additionally, this test is outperformed by the $ID_n$ test against both the $PM(0.5,3,5)$ and the $NB(15,0.75)$ distributions only by a single percentage point. The table further illustrates that $T_{n,L}^{(1)}$ is the second most powerful test against this class of distributions; this test outperforms each of the remaining tests against 4 of the remaining 7 alternatives. In summary, $T_{n,f_{n}}^{(1)}$ and $T_{n,L}^{(1)}$ either outperform, or produced empirical powers that are no less than $1\%$ inferior to the highest power achieved by the existing tests in 13 of the 14 cases considered.

Empirical powers obtained for samples of sizes 30 and 100 are available in Appendix B. The results for these sample sizes are also encouraging and generally exhibit similar patterns to those described above. As expected, the empirical powers increase with sample size. Interestingly, for samples of size 100, the $T_{n,f_{\widehat{\lambda}}}^{(1)}$ test is not outperformed by any other test against underdispersed alternatives. This phenomenon is surprising since it is not observed for samples of size 30 or 50. When increasing the sample size beyond 100 this test remains powerful, but it no longer outperforms all competitors.

\subsection{Practical applications}

We consider two practical examples in this section. The first pertains to the distribution of Sparrow nests, \cite{zar1999biostatistical} records the number of sparrow nests discovered on 40 one hectare plots. Table \ref{sparrow nests} shows the frequencies of the observed number of nests. As a second example, we consider the annual number of deaths due to horse kick in the Prussian army between 1875 and 1894. Table \ref{horsekicks} shows the observed frequency of these counts.

\begin{table}[!htbp] \centering 
  \caption{The number of sparrow nests found on 40 one hectare plots} 
  \label{sparrow nests} 
\begin{tabular}{@{\extracolsep{2pt}} lcccccccccccccccc} 
\\[-1.8ex]\hline 
\hline \\[-1.8ex] 
Count & \hspace{-0.3cm} $9$ & $22$ & $6$ & $2$ & $1$\\ 
Frequency & \hspace{-0.3cm} $0$ & $1$ & $2$ & $3$ & $4$\\
\hline \\[-1.8ex] 
\end{tabular} 
\end{table} 

\begin{table}[!htbp] \centering 
  \caption{The annual number of deaths due to horse kick in the Prussian army} 
  \label{horsekicks} 
\begin{tabular}{@{\extracolsep{2pt}} lcccccccccccccccc} 
\\[-1.8ex]\hline 
\hline \\[-1.8ex] 
Count & \hspace{-0.3cm} $3$ & $4$ & $5$ & $6$ & $7$ & $8$ & $9$ & $10$  & $11$ & $12$ & $13$ & $14$ & $15$ & $16$ & $17$ & $18$\\ 
Frequency & \hspace{-0.3cm} $1$ & $1$ & $2$ & $2$ & $1$ & $1$ & $2$ & $1$ & $3$ & $1$ & $0$ & $1$ & $2$ & $0$ & $1$ & $1$ \\
\hline \\[-1.8ex] 
\end{tabular} 
\end{table}

We test the hypotheses that the frequency distribution of the Sparrow nests, as well as that of the deaths due to horse kick are realised from a Poisson distributed. In each case, to estimate the p-values of the tests, a classical parametric bootstrap approach using 100 000 replications is employed, see \cite{gurtler2000recent}. The calculated test statistics and estimated p-values associated with each test are provided in Table \ref{results}, the results of the two examples are treated separately.

% Table created by stargazer v.5.2.3 by Marek Hlavac, Social Policy Institute. E-mail: marek.hlavac at gmail.com
% Date and time: Thu, Aug 04, 2022 - 10:24:06
\begin{table}[!htbp] \centering 
  \caption{The calculated test statistics and estimated p-values for the two practical examples} 
  \label{results} 
\begin{tabular}{@{\extracolsep{5pt}} lllll} 
\\[-1.8ex]\hline 
\hline \\[-1.8ex] 
& Sparrow nests & \hspace{2cm} & Horse kicks \\
Tests & Statistic & p-value & Statistic & p-value \\
\hline
$KS_n$ & $0.682$ & $0.037$ & $0.701$ & $0.095$ \vspace{0.1cm} \\ 
$CV_n$ & $0.000$ & $0.027$ & $0.000$ & $0.102$ \vspace{0.1cm} \\ 
$AD_n$ & $0.001$ & $0.054$ & $0.003$ & $0.017$ \vspace{0.1cm} \\ 
$KL_n$ & $1.364$ & $0.074$ & $5.094$ & $0.016$ \vspace{0.1cm} \\ 
$ID_n$ & $0.682$ & $0.050$ & $2.481$ & $0.013$ \vspace{0.1cm} \\ 
$T_{n,f_{\widehat{\lambda}}}^{(1)}$ & $0.377$ & $0.039$ & $0.705$ & $0.265$ \\ 
$T_{n,f_{n}}^{(1)}$ & $0.409$ & $0.092$ & $1.432$ & $0.142$ \\ 
$T_{n,L}^{(1)}$ & $0.574$ & $0.033$ & $1.776$ & $0.116$ \\ 
$T_{n,f_{\widehat{\lambda}}}^{(2)}$ & $0.155$ & $0.205$ & $1.179$ & $0.176$ \\ 
$T_{n,f_{n}}^{(2)}$ & $0.179$ & $0.268$ & $5.282$ & $0.182$ \\ 
$T_{n,L}^{(2)}$ & $0.223$ & $0.145$ & $2.673$ & $0.119$ \\ 
$T_{n,f_{\widehat{\lambda}}}^{(\infty)}$ & $0.184$ & $0.017$ & $0.075$ & $0.929$ \\ 
$T_{n,f_{n}}^{(\infty)}$ & $0.276$ & $0.064$ & $0.365$ & $0.437$ \\ 
$T_{n,L}^{(\infty)}$ & $0.324$ & $0.040$ & $1.000$ & $0.055$ \\ 
\hline \\[-1.8ex] 
\end{tabular} 
\end{table}

Consider the results pertaining to the frequency distribution of the sparrow nests. The p-values reported in Table \ref{results} indicate that, at a $10\%$ significance level, the null hypothesis is rejected by all of the tests considered, with the exception of $T_{n,f_{\widehat{\lambda}}}^{(2)}$ and $T_{n,f_{n}}^{(2)}$. As a result, we reject the null hypothesis and we conclude that the dataset is not realised from a Poisson distribution.

Next we consider the p-values obtained for the dataset relating to the deaths by horse kick in the Prussian army. For this dataset, 9 out of the 14 tests considered do not reject the Poisson assumption at a $10\%$ level of significance. In this case, we do not reject the Poisson assumption and we conclude that the observed dataset is sufficiently compatible with this assumption for the Poisson distribution to serve as an accurate model.

\section{Conclusions and recommendations}\label{four}

In this paper, we propose new goodness-of-fit tests for the Poisson distribution. These tests are related to a generalisation of the Poisson distribution known as the weighted Poisson. The probability mass function of a weighted Poisson random variable is obtained by multiplying that of a Poisson random variable with a weight function. In its most general form, the probability mass function of the weighted Poisson distribution can take on any form over the non-negative integers. As a result, we may choose a weight function so that the fitted probability mass function coincides exactly with the empirical mass function. We refer to the weight function for which this is the case as the empirical weight function. In the case of a given, large dataset which is realised from some Poisson distribution, the empirical weight function is expected to be close to $\mathbf{1}=\{1,1,\dots\}$. As a result, we may base a test for the Poisson distribution on some distance measure between the empirical weight function and the point $\mathbf{1}$.

We base tests on weighted $L_1$, $L_2$ and $L_{\infty}$ distances between the empirical weight function and $\mathbf{1}$. In each case, we use three different weight functions in the calculation of test statistics. These are the fitted Poisson mass function, the empirical mass function as well as a Laplace type kernel. The latter of these contains a tuning parameter, but the tests show remarkable insensitivity to the choice of the tuning parameter, meaning that we restrict our attention to a single choice. The three weighted distance measures, each calculated with reference to three different weight functions, result in a total of nine new goodness-of-fit tests for the Poisson distribution.

A Monte Carlo study is performed in order to compare the empirical powers of the newly proposed tests to that of existing tests. The Monte Carlo study comprises various sample sizes and employs a warp-speed bootstrap methodology in order to calculate empirical powers. We find that the newly proposed tests achieve the specified nominal significance level. Furthermore, these tests are highly competitive, generally outperforming the existing tests, against overdispersed alternatives. However, in certain settings, the newly proposed tests are not particularly powerful against underdispersed alternatives. The paper concludes with two practical examples demonstrating the use of goodness-of-fit tests in practice.

\titleformat{\section}{\large\bfseries}{\appendixname}{0.5em}{}

\appendix
\section{A: Proof of \eqref{infgenform} when $g(x)=f_{\widehat{\lambda}}(x)$}

The pmf of a $P(\lambda)$ random variable can be shown to be a non-increasing function in the case where $\lambda \leq 1$. On the other hand, if $\lambda>1$, then the pmf increases initially and decreases after attaining some maximum value. As a result, in order to calculate the final term in \eqref{infgenform}, we are required to determine the mode of a $P(\lambda)$ distribution. To this end, consider the following ratio:
\begin{eqnarray*}
\frac{f_\lambda(x)}{f_\lambda(x-1)}
&=& \left.\frac{\lambda^{x}e^{-\lambda}}{x!}\right/\frac{\lambda^{x-1}e^{-\lambda}}{(x-1)!} = \frac{\lambda}{x},
\end{eqnarray*}
which shows that
\begin{equation}
    f_{\lambda}(x) \ge f_{\lambda}(x-1) \Leftrightarrow \lambda \geq x. \label{ineq}
\end{equation}
As a result, $f_{\lambda}(x)$ is non-decreasing in $x$ for $x \le \lambda$ and non-increasing for $x \ge \lambda$. In the case where $\lambda$ is an integer, the $P(\lambda)$ distribution has two modes: $\lambda-1$ and $\lambda$. Turning our attention to non-integer $\lambda$, it follows from (\ref{ineq}) that the mode is $\lfloor \lambda \rfloor$, where $\lfloor x \rfloor$ denotes the integer part of $x$.

Combining the results for integer and non-integer $\lambda$, we know that $\lfloor \lambda \rfloor$ is a mode of the $P(\lambda)$ distribution. In the case of integer $\lambda$, this value corresponds to the larger of the two modes. However, we are interested in finding the smallest value of $x$ from which $g(x)=f_{\widehat{\lambda}}(x)$ is non-increasing so that we may calculate the final term in \eqref{infgenform}. Hence, the smallest mode is $\lceil \lambda \rceil - 1$, where $\lceil x \rceil$ denotes the ceiling of $x$. As a result, we have that $g(x)=f_{\widehat{\lambda}}(x)$ is non-increasing in $x \in \{\lceil \widehat{\lambda} \rceil - 1,\lceil \widehat{\lambda} \rceil,\dots \}$.

When computing \eqref{infgenform}, we are interested in determining whether $g(x)=f_{\widehat{\lambda}}(x)$ is non-increasing in $x \in \{m+1,m+2,\dots\}$. Note that 
\begin{eqnarray*}
    \lceil \widehat{\lambda} \rceil - 1 < \widehat{\lambda}
    = \frac{1}{n}\sum_{j=1}^{n} X_j
    \leq \frac{1}{n}\sum_{j=1}^{n} m
    < m + 1.
\end{eqnarray*}
Taking the above arguments into account, we have that $g(x)=f_{\widehat{\lambda}}(x)$ is non-increasing in $x \in \{m+1,m+2,\dots\}$. As a result, the final term in \eqref{infgenform} simplifies to
\begin{equation*}
    \underset{x \in \{m+1,m+2,\dots\}}{\textrm{max}}f_{\widehat{\lambda}}(x)=f_{\widehat{\lambda}}(m+1).
\end{equation*}

\section{B: Additional numerical results}

This appendix contains the empirical powers obtained for the sample sizes not discussed in the main text. Table \ref{n30} contains the results associated with samples of size 30, while Table \ref{n100} contains the results associated with samples of size of 100. As before, the tables contain the Fisher index of the alternative distributions used and, in order to ease comparison between the tests, the highest power against each alternative distribution is printed in bold.

\begin{landscape}
\begin{table}[!htbp] \centering 
  \caption{Empirical powers obtained for samples of size n = 30} 
  \label{n30} 
  \footnotesize
\begin{tabular}{@{\extracolsep{5pt}} lccccccccccccccc} 
\\[-1.8ex]\hline 
\hline \\[-1.8ex] 
Distribution & FI & $KS_n$ & $CV_n$ & $AD_n$ & $KL_n$ & $ID_n$ & $T_{n,f_{\widehat{\lambda}}}^{(1)}$ & $T_{n,f_{n}}^{(1)}$ & $T_{n,L}^{(1)}$ & $T_{n,f_{\widehat{\lambda}}}^{(2)}$ & $T_{n,f_{n}}^{(2)}$ & $T_{n,L}^{(2)}$ & $T_{n,f_{\widehat{\lambda}}}^{(\infty)}$ & $T_{n,f_{n}}^{(\infty)}$ & $T_{n,L}^{(\infty)}$ \\
\hline
$P(0.5)$ & $1.00$ & $5$ & $5$ & $5$ & $4$ & $5$ & $5$ & $5$ & $4$ & $5$ & $5$ & $5$ & $5$ & $5$ & $4$ \\ 
$P(1)$ & $1.00$ & $5$ & $5$ & $5$ & $5$ & $5$ & $5$ & $5$ & $5$ & $5$ & $5$ & $5$ & $5$ & $5$ & $4$ \\ 
$P(5)$ & $1.00$ & $5$ & $5$ & $5$ & $5$ & $5$ & $5$ & $5$ & $5$ & $5$ & $5$ & $5$ & $5$ & $5$ & $5$ \\ 
$P(10)$ & $1.00$ & $5$ & $5$ & $5$ & $5$ & $5$ & $5$ & $5$ & $5$ & $5$ & $5$ & $5$ & $5$ & $5$ & $5$ \\ 
$DU(4)$ & $1.00$ & $27$ & $32$ & $37$ & $34$ & $10$ & $\mathbf{38}$ & $18$ & $20$ & $14$ & $0$ & $14$ & $24$ & $17$ & $18$ \\ 
$B(5,0.25)$ & $0.75$ & $13$ & $12$ & $12$ & $11$ & $\mathbf{14}$ & $8$ & $1$ & $10$ & $1$ & $0$ & $2$ & $10$ & $3$ & $11$ \\ 
$B(5,0.2)$ & $0.80$ & $10$ & $\mathbf{10}$ & $9$ & $8$ & $\mathbf{10}$ & $7$ & $1$ & $8$ & $1$ & $0$ & $1$ & $9$ & $3$ & $9$ \\ 
$B(10,0.2)$ & $0.80$ & $7$ & $9$ & $8$ & $8$ & $\mathbf{10}$ & $6$ & $1$ & $5$ & $1$ & $1$ & $4$ & $8$ & $2$ & $5$ \\ 
$B(10,0.1)$ & $0.90$ & $6$ & $\mathbf{6}$ & $\mathbf{6}$ & $5$ & $\mathbf{6}$ & $\mathbf{6}$ & $2$ & $5$ & $2$ & $2$ & $2$ & $\mathbf{6}$ & $3$ & $5$ \\ 
$NB(9,0.9)$ & $1.11$ & $6$ & $6$ & $7$ & $7$ & $7$ & $6$ & $\mathbf{10}$ & $6$ & $\mathbf{10}$ & $\mathbf{10}$ & $\mathbf{10}$ & $5$ & $9$ & $6$ \\ 
$NB(45,0.9)$ & $1.11$ & $6$ & $6$ & $8$ & $8$ & $8$ & $6$ & $\mathbf{10}$ & $7$ & $9$ & $9$ & $7$ & $4$ & $9$ & $7$ \\ 
$PM(0.5,3,5)$ & $1.25$ & $10$ & $9$ & $14$ & $15$ & $15$ & $8$ & $\mathbf{16}$ & $10$ & $15$ & $14$ & $10$ & $4$ & $14$ & $10$ \\ 
$ZIP(0.9,3)$ & $1.30$ & $21$ & $14$ & $24$ & $21$ & $20$ & $19$ & $21$ & $43$ & $18$ & $10$ & $45$ & $9$ & $15$ & $\mathbf{48}$ \\ 
$PM(0.1,1,5)$ & $1.31$ & $13$ & $13$ & $22$ & $22$ & $21$ & $12$ & $23$ & $\mathbf{37}$ & $22$ & $15$ & $\mathbf{37}$ & $4$ & $18$ & $36$ \\ 
$NB(15,0.75)$ & $1.33$ & $12$ & $12$ & $19$ & $20$ & $21$ & $10$ & $\mathbf{23}$ & $13$ & $\mathbf{23}$ & $21$ & $13$ & $4$ & $20$ & $13$ \\ 
$NB(3,0.75)$ & $1.33$ & $14$ & $13$ & $18$ & $17$ & $19$ & $12$ & $\mathbf{25}$ & $16$ & $24$ & $23$ & $23$ & $7$ & $21$ & $14$ \\ 
$DU(6)$ & $1.33$ & $49$ & $53$ & $63$ & $62$ & $43$ & $52$ & $42$ & $63$ & $37$ & $3$ & $64$ & $23$ & $30$ & $\mathbf{65}$ \\ 
$NB(4,0.7)$ & $1.43$ & $18$ & $17$ & $25$ & $25$ & $27$ & $14$ & $\mathbf{31}$ & $17$ & $30$ & $28$ & $27$ & $8$ & $27$ & $17$ \\ 
$NB(2,2/3)$ & $1.50$ & $24$ & $22$ & $29$ & $28$ & $30$ & $20$ & $\mathbf{37}$ & $27$ & $36$ & $34$ & $34$ & $12$ & $32$ & $23$ \\ 
$NB(3,2/3)$ & $1.50$ & $23$ & $20$ & $29$ & $29$ & $32$ & $17$ & $\mathbf{36}$ & $21$ & $35$ & $32$ & $32$ & $10$ & $31$ & $20$ \\ 
$ZIP(0.8,3)$ & $1.60$ & $63$ & $45$ & $63$ & $56$ & $55$ & $55$ & $54$ & $78$ & $48$ & $21$ & $\mathbf{79}$ & $36$ & $40$ & $\mathbf{79}$ \\ 
$PM(0.2,1,5)$ & $1.61$ & $38$ & $35$ & $53$ & $53$ & $53$ & $30$ & $47$ & $\mathbf{61}$ & $45$ & $28$ & $60$ & $7$ & $36$ & $60$ \\
$NB(1,0.5)$ & $2.00$ & $54$ & $53$ & $60$ & $57$ & $59$ & $45$ & $\mathbf{65}$ & $58$ & $63$ & $59$ & $62$ & $31$ & $57$ & $48$ \\ 
\hline \\[-1.8ex] 
\end{tabular} 
\end{table} 
\end{landscape}

\begin{landscape}
\begin{table}[!htbp] \centering 
  \caption{Empirical powers obtained for samples of size n = 100} 
  \label{n100}
  \footnotesize
\begin{tabular}{@{\extracolsep{5pt}} lccccccccccccccc} 
\\[-1.8ex]\hline 
\hline \\[-1.8ex] 
Distribution & FI & $KS_n$ & $CV_n$ & $AD_n$ & $KL_n$ & $ID_n$ & $T_{n,f_{\widehat{\lambda}}}^{(1)}$ & $T_{n,f_{n}}^{(1)}$ & $T_{n,L}^{(1)}$ & $T_{n,f_{\widehat{\lambda}}}^{(2)}$ & $T_{n,f_{n}}^{(2)}$ & $T_{n,L}^{(2)}$ & $T_{n,f_{\widehat{\lambda}}}^{(\infty)}$ & $T_{n,f_{n}}^{(\infty)}$ & $T_{n,L}^{(\infty)}$ \\
\hline
$P(0.5)$ & $1.00$ & $5$ & $5$ & $5$ & $5$ & $5$ & $5$ & $5$ & $5$ & $5$ & $5$ & $5$ & $5$ & $5$ & $5$ \\ 
$P(1)$ & $1.00$ & $5$ & $5$ & $5$ & $5$ & $5$ & $5$ & $5$ & $5$ & $5$ & $5$ & $5$ & $5$ & $5$ & $5$ \\ 
$P(5)$ & $1.00$ & $5$ & $5$ & $5$ & $5$ & $5$ & $5$ & $5$ & $5$ & $5$ & $5$ & $5$ & $5$ & $5$ & $5$ \\ 
$P(10)$ & $1.00$ & $5$ & $5$ & $5$ & $5$ & $5$ & $5$ & $5$ & $5$ & $5$ & $5$ & $5$ & $5$ & $5$ & $5$ \\ 
$DU(4)$ & $1.00$ & $80$ & $89$ & $\mathbf{96}$ & $94$ & $45$ & $93$ & $84$ & $65$ & $81$ & $0$ & $58$ & $79$ & $79$ & $57$ \\ 
$B(5,0.25)$ & $0.75$ & $33$ & $32$ & $37$ & $37$ & $\mathbf{44}$ & $24$ & $6$ & $29$ & $2$ & $0$ & $2$ & $20$ & $10$ & $33$ \\ 
$B(5,0.2)$ & $0.80$ & $24$ & $22$ & $24$ & $23$ & $\mathbf{27}$ & $18$ & $5$ & $20$ & $1$ & $0$ & $0$ & $15$ & $9$ & $24$ \\ 
$B(10,0.2)$ & $0.80$ & $17$ & $19$ & $22$ & $23$ & $\mathbf{29}$ & $10$ & $1$ & $12$ & $1$ & $0$ & $9$ & $12$ & $5$ & $15$ \\ 
$B(10,0.1)$ & $0.90$ & $9$ & $8$ & $8$ & $8$ & $\mathbf{9}$ & $7$ & $2$ & $7$ & $1$ & $1$ & $1$ & $7$ & $4$ & $8$ \\ 
$NB(9,0.9)$ & $1.11$ & $10$ & $9$ & $10$ & $11$ & $11$ & $8$ & $14$ & $10$ & $\mathbf{15}$ & $14$ & $13$ & $7$ & $12$ & $9$ \\ 
$NB(45,0.9)$ & $1.11$ & $8$ & $8$ & $11$ & $11$ & $12$ & $7$ & $\mathbf{13}$ & $9$ & $12$ & $11$ & $9$ & $5$ & $11$ & $9$ \\ 
$PM(0.5,3,5)$ & $1.25$ & $20$ & $21$ & $31$ & $32$ & $\mathbf{35}$ & $13$ & $26$ & $19$ & $24$ & $17$ & $19$ & $7$ & $18$ & $19$ \\
$ZIP(0.9,3)$ & $1.30$ & $61$ & $39$ & $62$ & $55$ & $54$ & $52$ & $55$ & $\mathbf{86}$ & $44$ & $12$ & $\mathbf{86}$ & $34$ & $45$ & $\mathbf{86}$ \\ 
$PM(0.1,1,5)$ & $1.31$ & $35$ & $34$ & $53$ & $54$ & $53$ & $28$ & $51$ & $\mathbf{77}$ & $46$ & $18$ & $76$ & $10$ & $38$ & $75$ \\ 
$NB(15,0.75)$ & $1.33$ & $27$ & $29$ & $44$ & $46$ & $\mathbf{50}$ & $17$ & $41$ & $21$ & $39$ & $30$ & $21$ & $7$ & $30$ & $20$ \\ 
$NB(3,0.75)$ & $1.33$ & $37$ & $35$ & $42$ & $43$ & $46$ & $30$ & $\mathbf{48}$ & $41$ & $44$ & $38$ & $38$ & $20$ & $35$ & $38$ \\ 
$DU(6)$ & $1.33$ & $97$ & $98$ & $\mathbf{100}$ & $99$ & $94$ & $99$ & $98$ & $97$ & $97$ & $16$ & $97$ & $83$ & $91$ & $97$ \\ 
$NB(4,0.7)$ & $1.43$ & $47$ & $46$ & $58$ & $60$ & $\mathbf{63}$ & $35$ & $61$ & $43$ & $56$ & $45$ & $57$ & $22$ & $46$ & $45$ \\ 
$NB(2,2/3)$ & $1.50$ & $63$ & $61$ & $68$ & $68$ & $70$ & $54$ & $\mathbf{71}$ & $67$ & $66$ & $56$ & $58$ & $39$ & $55$ & $63$ \\ 
$NB(3,2/3)$ & $1.50$ & $60$ & $58$ & $69$ & $69$ & $\mathbf{73}$ & $47$ & $70$ & $57$ & $66$ & $54$ & $67$ & $32$ & $54$ & $57$ \\ 
$ZIP(0.8,3)$ & $1.60$ & $99$ & $94$ & $99$ & $98$ & $98$ & $98$ & $98$ & $\mathbf{100}$ & $96$ & $35$ & $\mathbf{100}$ & $95$ & $97$ & $\mathbf{100}$ \\ 
$PM(0.2,1,5)$ & $1.61$ & $88$ & $85$ & $96$ & $96$ & $96$ & $78$ & $92$ & $\mathbf{98}$ & $88$ & $46$ & $97$ & $38$ & $79$ & $97$ \\ 
$NB(1,0.5)$ & $2.00$ & $96$ & $96$ & $\mathbf{97}$ & $96$ & $\mathbf{97}$ & $92$ & $\mathbf{97}$ & $\mathbf{97}$ & $95$ & $89$ & $92$ & $84$ & $90$ & $94$ \\ 
\hline \\[-1.8ex] 
\end{tabular} 
\end{table} 
\end{landscape}

\bibliographystyle{apalike}
\bibliography{refs}

\end{document}